\documentclass[11pt]{article}

\usepackage{amsfonts, amssymb}
\usepackage{amsmath}
\usepackage{amsthm}

\newtheorem{theorem}{\bf Theorem}[section]

\newtheorem{corollary}{\bf Corollary}[section]
\newtheorem{definition}{\bf Definition}[section]
\newtheorem{lemma}[theorem]{Lemma}

\begin{document}

\begin{center}
{\LARGE\bf Global branching of solutions to ODEs and integrability}
\\
\vskip 2 mm
{\Large Rod Halburd}\footnote{r.halburd@ucl.ac.uk}
\\\vskip 2 mm
Department of Mathematics, University College London, 
\\
Gower Street, London WC1E 6BT, UK
\end{center}

\begin{abstract}
We consider a natural generalisation of the Painlev\'e property and use it to identify the known integrable cases of the Lane-Emden equation with a real positive index.
We classify certain first-order ordinary differential equations with this property and find necessary conditions for a large family of second-order equations.
We consider ODEs  such that, given any simply connected domain $\Omega$ not containing fixed singularities of the equation, the Riemann surface of any solution obtained by analytic continuation along curves in $\Omega$ has a finite number of sheets over $\Omega$.
\end{abstract}

\section{Introduction}
The Painlev\'e property is frequently used to detect 
integrable differential equations.  Simple methods can be applied to determine strong necessary conditions for an ordinary differential equation to have this property. It is particularly useful for finding special values of parameters or special forms of otherwise arbitrary functions of the independent variable for which an equation is in some sense integrable.  An ODE is said to possess the Painlev\'e property if all solutions are single-valued about all movable singularities.  Sometimes a more restrictive version of this property is used, which we refer to as the {\em strong Painlev\'e property}, namely that all movable singularities of all solutions are poles.  A precise definition of a movable singularity can be found in \cite{kimura:56,murata:88}, but roughly speaking, a singularity of a solution is movable if its location moves as we vary the initial conditions, as opposed to fixed singularities, which are values of the independent variable at which the ODE is singular in some sense.

The Painlev\'e property is very sensitive to changes in dependent variables, which means in practice that one wants to find a preferred set of variables with which to work.
Certainly there are many tricks one can use to guess appropriate coordinate changes, but these are usually {\it ad hoc} and not particularly useful when trying to prove theorems of a very general nature.

A natural idea is to allow for some mild branching around movable singularities.
In \cite{ramanidg:82}, Ramani, Dorizzi and Grammaticos
considered the so-called {\em weak Painlev\'e property} in which solutions are permitted to have series expansions of the form
\[
\sum_{j=0}^\infty a_j(z-z_0)^{(j+p)/q},
\]
about movable singularities $z_0$.
However, in a later paper \cite{ranadarg:85}, the authors of \cite{ramanidg:82}, together with Ranada, showed that this property alone is too weak to single out integrable equations.  In fact, Painlev\'e \cite{painleve:88} showed  that any first-order equation of the form
$F(z;w,w')=0$ has this property, where $F$ is a polynomial in $w$ and $w'$ with coefficients that are analytic in $z$ on some open set.

Smith \cite{smith:53} studied a class of equations and showed that all movable singularities of all solutions that can be reached by analytic extension along a curve of finite length must be algebraic.  Shimomura \cite{shimomura:07,shimomura:08} called this property the {\em quasi-Painlev\'e property} and proved that certain higher degree generalisations of the first and second Painlev\'e equations have this property. Filipuk and author \cite{flilipukh:09a,flilipukh:09b,flilipukh:09c} and Kecker \cite{kecker:12,kecker:16} considered much more general classes of second-order ODEs with arbitrary coefficients and showed that, provided certain formal series solutions exist, then the equations have the quasi-Painlev\'e property.

Despite the fact that in a neighbourhood of such a singularity the solution is defined on a Riemann surface with a finite number of sheets, generically the global singularity structure is much more complicated and typically the maximal analytic extension of the solution requires a Riemann surface with an infinite number of sheets over $\mathbb{C}$.
For example, consider an equation of the form $(w')^2=P(w)$, where $P$ is a polynomial of degree $2g+1$ or $2g+2$ with no repeated roots and $g>0$.  In the generic case, the solutions of such an equation will have $2g$ periods that are linearly independent over $\mathbb{Z}$ and are pairwise linearly independent over $\mathbb{R}$.  So generically, when $P$ has degree at least five, in which case $g>1$, there are at least three periods independent in the above sense, in which case $w$ cannot be meromorphic.  Moreover, the maximal analytic continuation of $w$ is not defined on a Riemann surface with only a finite number of sheets.  In fact, the solution must be {\em densely branched} in the sense that we can choose any $z_0\in\mathbb{C}$, any $w_0\in\mathbb{C}\setminus\{0\}$ and any $\epsilon>0$ and there will be a closed path $\Gamma\subset\mathbb{C}$ starting from $z_0$ such that the solution $w$ satisfying the initial condition $w(z_0)=w_0$ can be analytically continued along $\Gamma$ such that when $\Gamma$ returns to $z_0$, the value $w_1$ of the analytic continuation of $w$ at $z=z_0$ satisfies $|w_0-w_1|<\epsilon$.

With such examples in mind, Kruskal and Clarkson \cite{kruskalc:92} introduced the {\em poly-Painlev\'e} property. 
An ODE is said to possess the poly-Painlev\'e property if no solution exhibits dense branching.  
Note that this definition does not distinguish between fixed and movable singularities.  In particular, Kruskal and Clarkson consider the linear ODE
\[
	\frac{{\rm d}w}{{\rm d}z}=\sum_{j=1}^m\frac{a_j}{z-z_j},
\]
where $a_j$ and $z_j$ are complex constants, $j=1,\ldots,m$.  This equation does not have any movable singularities but for $m>2$ and for generic complex numbers $a_1,\ldots,a_m$, the general solution exhibits dense branching because the solution is only determined up to an additive factor of the form 
$2\pi{\rm i}(n_1a_1+\cdots+n_ma_m)$, where $n_j\in\mathbb{Z}$, $j=1,\ldots,m$,
and so the equation does not possess the poly-Painlev\'e property.

In this paper we study a variation on this theme, which appears to have first been considered by Painlev\'e himself, still demanding that the global branching structure of solutions not be too complicated.  First, we choose to ignore branching that comes from fixed singularities, so that in particular we do not exclude linear equations such as the hypergeometric equation.  We then require that any branching that arises from analytic continuation along paths that do not enclose fixed singularities is only finite, in the sense that the ``multi-valued'' solution can only have a finite number of values at any point.

\begin{definition}
\label{def}
An ODE is said to possess the \underline{algebro-Painlev\'e property} if, given any solution $w$ and a point $z_0$ in the domain of analyticity of $w$, there are only finitely many values that can be taken at $z_0$ by any analytic continuation of $w$ around closed curves not enclosing a fixed singularity of the equation.
\end{definition}
\noindent
If any of the coefficients of the equation are themselves branched, then the curves and their interiors should be considered on the appropriate Riemann surface.

Consider the equation
\begin{equation}
\label{non-auto}
	\frac{{\rm d} w}{{\rm d}z}=R(z,w),
\end{equation}
where $R$ is a rational function of $z$ and $w$.
Painlev\'e himself showed that if equation (\ref{non-auto}) possesses the algebro-Painlev\'e property (as given in Definition \ref{def}) then it can be transformed to a Riccati equation
\begin{equation}
\label{riccati}
	\frac{{\rm d} u}{{\rm d}z}= a(z)u^2+b(z)u+c(z),
\end{equation}
where $a$, $b$ and $c$ are rational functions.

Explicitly, Painlev\'e  \cite{painleve:88} proved the following.
\begin{theorem}
[Painlev\'e]
\label{painleve-thm}
If any solution of equation (\ref{non-auto}), except at most a denumerable set of solutions, has exactly $n$ $(<\infty)$ branches which are mutually permutable around its movable branch points, then equation (\ref{non-auto}) can be reduced to a Riccati equation (\ref{riccati}) by a transformation of the form
\begin{equation}
\label{transformation}
	u=\frac{\alpha_0(z)+\alpha_1(z)w+\cdots+\alpha_{n-1}(z)w^{n-1}+w^n}{\beta_0(z)+\beta_1(z)w+\cdots+\beta_{n-1}(z)w^{n-1}+w^n}
\end{equation}
where $\alpha_0,\ldots\alpha_{n-1};\beta_0,\ldots,\beta_{n-1};a,b$ and $c$ are all rational functions of $z$.
\end{theorem}

A related result was stated by Malmquist \cite{malmquist:13}.
\begin{theorem}
[Malmquist]
\label{malmquist-thm}
If equation (\ref{non-auto}) admits at least one solution which is a finitely many-valued transcendental function and if $n$ denotes the number of branches which permute mutually around its movable branch points, then equation (\ref{non-auto}) can be reduced to equation (\ref{riccati}) by (\ref{transformation}).
\end{theorem}

T.~Kimura \cite{kimura:66} later proved generalisations of Theorem \ref{painleve-thm} and Theorem \ref{malmquist-thm}. One difficulty in the application of these theorems to identify and solve integrable first-order equations is that the number $n$ of sheets is not obvious from the local branching behaviour alone. 
The difficulty in deducing global branching from local branching of solutions to an ODE is apparent from the simple example of the constant coefficient Abel equation of the first kind:
\begin{equation}
\label{abel}
	u'(z)=u(z)\{u(z)-a\}\{u(z)-b\},\qquad a,b\in\mathbb{C},
\end{equation}
where we will assume that $a$, $b$ and $0$ are pairwise distinct.  Each singularity $z_0$ of equation (\ref{abel}) is a square-root branch point with leading-order behaviour proportional to $1/\sqrt{z-z_0}$. The general solution of equation (\ref{abel}) is given implicitly by
\[
	\left(
		\frac{u-a}u
	\right)^b
	\left(
		\frac{u}{u-b}
	\right)^a
	=
	\kappa {\rm e}^{ab(a-b)z},
\]
for some constant $\kappa$.
If $a$ and $b$ are both rational numbers, then each solution $u(z)$ is finitely-valued, but we can choose $a$ and $b$ to make the number of values arbitrarily high.  
For example, we see that if $a=1$ and $b=n$, where $n>1$ is an integer, then $u$ is the root of a polynomial equation of degree $n$:
\[
	(u-1)^{n}-\kappa{\rm e}^{n(n-1)z}u^{n-1}(u-n)=0.
\]

In \cite{halburdk:14}, Kecker and the author also defined a property called the {\em algebro-Painlev\'e property}.
In this paper, we will refer to the definition in \cite{halburdk:14} as the {\em strong algebro-Painlev\'e property}.
Let $F$ be the set of fixed singularities of some ODE and let $M$ be the set of meromorphic functions over $\mathbb{C}\setminus F$. The equation is said to have the strong algebro-Painlev\'e property if all solutions are algebraic over $M$.

As one application, we will use the algebro-Painlev\'e property (as given in Definition \ref{def}) to identify integrable cases of stellar models with polytropic equations of state.  Such models are described by the Lane-Emden equation
of index $n$,
\begin{equation}
\label{laneemden}
	\xi^{-2}\frac{{\rm d}\ }{{\rm d}\xi}\left(\xi^2\frac{{\rm d}\theta}{{\rm d}\xi}\right)=-\theta^n,
\end{equation}
subject to the initial conditions 
\begin{equation}
\label{BCs}
	\theta(0)=1\quad\mbox{and}\quad\theta'(0)=0.
\end{equation}
It should be noted that $n$ is any non-negative real number, not necessarily an integer.
It is well known (see, e.g. \cite{chandrasekhar:39}) that this initial value problem can be solved in closed form in the cases $n=0$, $n=1$ and $n=5$.  Also, the general solution to equation (\ref{laneemden}) can be written down explicitly for each of these three values of $n$.

In section \ref{sect-first-order} we will give necessary and sufficient conditions for certain first-order ODEs to possess the algebro-Painlev\'e property.  As well as being of independent interest, these results will be used in our analysis of second-order equations in subsequent sections.

In section \ref{sec:laneemden} we will prove the following.
\begin{theorem}
\label{lane-emden-thm}
Let $n$ be real and non-negative.  Then the Lane-Emden equation (\ref{laneemden}) of non-negative index $n\in\mathbb{R}$ possesses the algebro-Painlev\'e property if and only if $n=0$, $n=1$ or $n=5$.
\end{theorem}

Section \ref{sec:general} concerns the more general equation
\begin{equation}
\label{general-ode}
	\frac{{\rm d}^2w}{{\rm d}z^2}=L(z,w)\left(\frac{{\rm d}w}{{\rm d}z}\right)^2+M(z,w)\frac{{\rm d}w}{{\rm d}z}+N(z,w),
\end{equation}
where $L(z,w)$, $M(z,w)$ and $N(z,w)$ are rational functions of $w$ with coefficients that are analytic functions of $z$ on some common domain $\Omega$.
We will prove several results providing necessary conditions that constrain the forms of $L(z,w)$, $M(z,w)$ and $N(z,w)$ such that equation (\ref{general-ode}) possesses the algebro-Painlev\'e property.

\section{First-order equations}
\label{sect-first-order}

Throughout this section we will only consider autonomous ODEs. In this case, the algebro-Painlev\'e property is equivalent to the requirement that the maximal analytic extension of each solution is finitely-many valued.
We will prove some results about first-order ODEs that are of interest in their own right, but they will also play an important role in our analysis of the Lane-Emden equation and other second-order equations in later sections of this paper.

Before we move on to more general equations, we will make a few remarks about steepest ascent curves of holomorphic functions.
The steepest ascent curve of a conformal mapping $f$ at $z_0\in\mathbb{C}$ is the (directed) curve through $z_0$ along which the rate of change of $|f(z)|$ is the greatest.  Each such curve is an integral curve of the gradient vector field of $|f(z)|$.  The argument of $f(z)$ is a constant along the steepest ascent curve through $z_0$.  This means that if $f(z_0)\ne 0$ then we can parametrise the steepest ascent curve $\gamma(t):[0,1)\to\mathbb{C}$ such that
\[
	f\circ\gamma(t)=\sigma(t)f(z_0),
\]
where $\sigma(t):[0,1)\to \mathbb{R}_{>0}$ is increasing and $\sigma(0)=1$.

The following lemma will be useful in this section.

\begin{lemma}
\label{ascent}
Let $f$ be such that $f'(z)=(z-a)^{-r}g(z)$, where $r\ge 1$ is a rational number, $g$ is holomorphic on the disc $D=\{z\in\mathbb{C}:0<|z-a|<\epsilon\}$ of radius $\epsilon>0$ centred at $a$, and $g(a)\ne 0$.
If $r>1$ then given any $\xi\in\mathbb{C}\setminus\{0\}$ there is a curve $\gamma:[0,1)\to U$ such that 
$f\circ\gamma(t)=\rho(t)\xi$, where $\rho(t)$ is a positive increasing function of $t$ such that $\rho(t)\to\infty$ as $t\to 1^-$.
If $r=1$, the same conclusion holds subject to the additional condition that $\mbox{Re}(\xi/g(a))\ne 0$.
\end{lemma}
A similar result holds when $a=\infty$.
The basic idea of the proof is that, for $r>1$, $f(z)$ is well approximated by $\frac{g(a)}{1-r}\frac{1}{(z-a)^{r-1}}$ and the curves of constant argument are approximately straight lines approaching $a$. When $r=1$, $f(z)$ is well approximated by $g(a)\log(z-a)$.  Let $z=a+r{\rm e}^{{\rm i}\theta}$. Then demanding that $g(a)\log(z(t)-a)=\rho(t)\xi$, for some $\xi$ gives
\[
	\log r(t)+{\rm i}\theta(t)=(\mu_0+{\rm i}\nu_0)\sigma(t),
\]
where $\xi/g(a)=\mu_0+{\rm i}\nu_0$. When $\mu_0\ne 0$, we see that the steepest ascent curve is given by
$\theta=-\frac{\nu_0}{\mu_0}\log(1/r)$, which spirals infinitely many times around $a$ if $\nu_0\ne 0$.

Now we consider the equation
\begin{equation}
\label{first-rational}
	\frac{{\rm d}u}{{\rm d}z}=R(u),
\end{equation}
where $R$ is a rational function.
Observe that (\ref{first-rational}) does not have the algebro-Painlev\'e property if and only if there exists $z_0\in\mathbb{C}$ and an infinite number of pairwise distinct complex numbers
$u_0,u_1,\ldots$ and an infinite number of curves $\gamma_1,\gamma_2,\ldots$, where each $\gamma_j$ is a curve from $u_0$ to $u_j$, such that for all $j=1,2,\ldots\ $,
\begin{equation}
\label{zero-int}
	\int_{\gamma_j}\frac{{\rm d}v}{R(v)}=0.
\end{equation}

Let $a_1,\ldots,a_n$ be the zeros of $R$ and let $\alpha_j$ be the residue of $1/R(v)$ at $v=a_j$.  Let $u_0$ be a complex number that is neither a zero nor a pole of $R$. Then we can solve equation (\ref{first-rational}) with the initial condition $u(z_0)=u_0$ by separating variables in equation  (\ref{first-rational}) and integrating along a curve $\gamma$ in the complex $v$ plane from $u_0$ to $u$,
giving
\begin{equation}
\label{Zdef}
	Z(u):=Q(u)+\sum_{j=1}^n\alpha_j\int_{\gamma}\frac{{\rm d}u}{u-a_j}=z-z_0,
\end{equation}
where $Q(u)$ is a rational function satisfying $Q(u_0)=0$.
Clearly if all residues $\alpha_j$ are zero, then $u$ is an algebraic function of $z$.  From now on we suppose that $\alpha_k\ne 0$ for some $k\in\{1,\ldots,n\}$.

From Lemma \ref{ascent}, we see that if the rational function $Q(u)$ is not a constant, then there exists a (steepest ascent) path $\gamma(t)$ in the complex $u$-plane such that $Z(\gamma(t))=2\pi{\rm i}\alpha_k\sigma(t)$, where $\sigma(t)$ is an increasing real-valued function of $t$ on $[0,1)$ and $\sigma(t)\mapsto \infty$ as $t$ tends to $1$ from below and $\gamma(t)$  tends either to infinity or to $u_l$ for some $l\in\{1,\ldots,n\}$.  Without loss of generality, take $u_0=\gamma(0)$.  For each $j=0,1,2,\ldots$, define $t_j\in[0,1)$ such that
$\sigma(t_j)=j$ and let $u_j=\gamma(t_j)$.  Let $\gamma_j$ be the path from $u_0$ along $\gamma$ to the point $u_j$ followed by a closed path from $u_j$ back to itself that loops $j$ times in the negative direction around $a_j$.  For each such path, equation (\ref{zero-int}) holds.  Hence if equation (\ref{first-rational}) has the algebro-Painlev\'e property, then $Q$ is a constant.  From now on we assume that $Q\equiv 0$.

Suppose that there is a number $\xi\in\mathbb{C}\setminus\{0\}$ such that for all $j\in\{1,\ldots,n\}$, $\alpha_j = \beta_j\xi$ where $\beta_j\in\mathbb{Q}$.  Then it follows from equation (\ref{Zdef}) that $u$ is an algebraic function of 
${\rm e}^{z/\xi}$, and hence finitely many-valued.  
It remains to consider the case in which, possibly after re-indexing, we have $n>1$, $\alpha_1\ne 0$ and 
$\alpha_2=(\mu+{\rm i}\nu)\alpha_1$,
where $\mu$ and $\nu$ are real and $\mu+{\rm i}\nu\not\in{\mathbb{Q}}$.
 Let $\gamma(t)$, $t\in[0,1]$ be a curve in the complex $u$-plane such that $\gamma(1)=a_1$ and $Z(\gamma(t))=2\pi\rho(t)\alpha_1$, where $\rho(t)$ is an increasing real-valued function, $\rho(t)\to\infty$ as $t$ tends to $1$ from below. 
If $\mu=0$ then for all $j=0,1,2,\ldots$, define $t_j$ by $\rho(t_j)=j$ and let $u_j=\gamma(t_j)$.  Define the curve $\gamma_j$ to be the curve from $u_0$ to $u_j$ along $\gamma$, followed by a closed curve from $u_j$ back to $u_j$ that loops around $a_2$ $j$ times in the positive direction.  For each such path, equation (\ref{zero-int}) holds and equation (\ref{first-rational}) does not have the algebro-Painlev\'e property.

Now assume that $\mu\ne 0$.  Let $\epsilon_l$ be a sequence of positive numbers such that $\epsilon_l\mapsto 0$ as $l\mapsto\infty$ and choose integers $m_l$ and $n_l$, $n_l\mapsto\infty$, such that $|m_l+n_l\mu|<\epsilon_l/(2\pi\alpha_1)$. Let $t_j$ be defined by $\rho(t_j)=j$, let $u_j=\gamma(t_j)$ and observe that $u_j\to a_1$.  Finally, let $\gamma_j$ be the curve from $u_0$ to $u_j$ along $\gamma$ followed by a closed curve from $u_j$ back to $u_j$ that loops around $a_2$ $j$ times in the positive direction. Then $u(z_j)=u_j$, where $z_j=z_0+2\pi{\rm i}(m_j+n_j\mu)\alpha_1$.  So as $j\to\infty$, $z_j\mapsto z_0$ and $u(z_j)=u_j\to a_1$.  It follows that if the maximal analytic continuation of $u$ were defined on a Riemann surface with a finite number of sheets, then there would be a subsequence of $(z_j)$ converging to $a_1$ on one of them, so $u$ would be identically $a_1$.  Therefore we conclude that $u$ is infinitely many-valued.

We have proved the following.

\begin{theorem}
\label{rat-thm}
Let $R$ be a rational function.  The maximal analytic extension of every solution of
\[
	\frac{{\rm d}u}{{\rm d}z}=R(u)
\]
is finitely-valued if and only if one of the following holds:
\begin{enumerate}
	\item $R\equiv 0$; or
	\item at every zero of $R$, the residue of $1/R$ is zero; or
	\item $1/R$ has the form
		\[
			\frac 1{R(v)}=\xi\sum_{j=1}^n\frac{\beta_j}{v-a_j},
		\]
\end{enumerate}
for some integer $n>0$, $\xi\in\mathbb{C}\setminus\{0\}$, $a_j\in\mathbb{C}$ and $\beta_j\in \mathbb{Q}$ for all $j\in\{1,\ldots,n\}$.
\end{theorem}

In section 4 we will need the following.
\begin{corollary}
\label{deg1-cor}
Consider the equation
\begin{equation}
\label{fod-ode}
	\frac{{\rm d}u}{{\rm d}z}
	=
	\frac{u^s-1}{u^{r}},
\end{equation}
where $r$ and $s$ are rational numbers, with $s> 0$.  
All maximally-extended solutions of equation (\ref{fod-ode}) are finitely-many valued if and only if $(r+1)/s=0$, $1/2$ or $1$.
For each of these cases, the non-constant solutions are given below.

\begin{itemize}
\item
When $r=-1$,
\[
	u(z)=	\left\{
			1-\exp\left(s[z-z_0]\right)
		\right\}^{1/s}.
\]
\item
When $(r+1)/s=1/2$,
\[
	u(z)=	\left\{
			\coth\left(\frac s2\left[z-z_0\right]\right)
		\right\}^{2/s}.
\]
\item
When $(r+1)/s=1$,
\[
	u(z)=	\left\{
			1+\exp\left(s[z-z_0]\right)
		\right\}^{1/s}.
\]
\end{itemize}
\end{corollary}

Notice that if $r\ne -1$ then the transformation $u\mapsto u^{1/(1+r)}$, $z\mapsto z/(1+r)$ maps equations (\ref{fod-ode})  to an equation of the same form with $r=0$ and $s$ replaced by $s/(1+r)$.

\vskip 3mm

\noindent{\bf Proof.}

Let $c$ be the lowest common denominator of $r$ and $s$.  Then there are integers $a$ and $b$ such that $r=(a-c)/c$ and $s=b/c$.  
On making the change of variables $v(Z)=u(z)^{1/c}$, $Z=z/c$ in equation (\ref{fod-ode}) and separating variables, we obtain up to an additive integration constant,
\begin{equation}
\label{ab1-sep}
	Z
	=
	\int\frac{v^{a-1}}{v^b-1}{\rm d}v.
\end{equation}
If $a\ge 1$, then there are integers $m\ge 0$ and $0\le k\le b-1$ such that $a=mb+k+1$. Then
\begin{equation}
\begin{split}
	\frac{v^{a-1}}{v^b-1}
	&=
	v^{k}\left(1+v^b+v^{2b}+\cdots+v^{(m-1)b}\right)
	+
	\frac{v^{k}}{v^b-1}
	\\
	&=P(v)
	+
	\displaystyle\frac 1b\sum_{j=0}^{b-1}\frac{\omega^{j(k+1)}}{v-\omega^j},
\end{split}
\end{equation}
where $P$ is a polynomial and $\omega={\rm e}^{2\pi{\rm i}/b}$ is the fundamental $b$th root of unity.
It follows from Theorem \ref{rat-thm} that $P=0$ and $\omega^{k+1}\in\mathbb{Q}$.
The fact that $P=0$ means that $m=0$, so $1\le a=k+1\le b$.
Since $\omega^{k+1}$ is rational, we have $\omega^{k+1}=\pm1$. So either $k=b-1$ or $b$ is even and $k=(b-2)/2$.
That is, either $a=b$ or $b$ is even and $a=b/2$.
When $a=b$, equation (\ref{ab1-sep}) gives $v^b=1+\exp\left(b[Z-Z_0]\right)$, where $Z_0$ is a constant.
When $b=2d$ and $a=d$, equation (\ref{ab1-sep}) gives $v^d=-\coth\left(d[Z-Z_0]\right)$.

If $a\le 0$, then there are integers $m\ge 0$ and $0\le k\le b-1$ such that $a=-(mb+k)$. Then
\begin{equation}
\begin{split}
	\frac{v^{a-1}}{v^b-1}
	&=
	-v^{-(k+1)}\left(1+v^{-b}+v^{-2b}+\cdots+v^{-mb}\right)
	+
	\frac{v^{b-k-1}}{v^b-1}
	\\
	&=v^{-b-k-1}Q\left(v^{-1}\right)
	-\frac1{v^{k+1}}
	+
	\displaystyle\frac 1b\sum_{j=0}^{b-1}\frac{\omega^{-jk}}{v-\omega^j},
\end{split}
\end{equation}
where $Q$ is a polynomial.  From Theorem \ref{rat-thm}, we see that any pole at $v=0$ must be simple, so $m=0=k$, which gives $a=0$. 
Equation (\ref{ab1-sep}) can now be integrated to get $v^{-b}=1+\exp(b[Z-Z_0])$.
\hfill$\Box$

\vskip 3 mm

Next we consider the equation
\begin{equation}
\label{fo-ode}
	\left(\frac{{\rm d}u}{{\rm d}z}\right)^2
	=
	\frac{u^s-1}{u^{r}},
\end{equation}
where $r$ and $s$ are rational numbers, $s\ne 0$.
If $s<0$, then $\tilde u(\tilde z):=u(z)$ solves
\[
	\left(\frac{{\rm d}\tilde u}{{\rm d}\tilde z}\right)^2
	=
	\frac{{\tilde u}^{\tilde s}-1}{{\tilde u}^{\tilde r}},
\]
where $\tilde z={\rm i}z$, $\tilde s=-s$ and $\tilde r=r-s$.
So without loss of generality, we assume that $s>0$ in equation (\ref{fo-ode}).  

Let $c>0$ be the lowest common denominator of $r/2$ and $s$.  Then there are integers $a$ and $b$ such that $r=2(a-c)/c$ and $s=b/c$.  On making the change of variables $v(Z)=u(z)^{1/c}$, $Z=z/c$ in equation (\ref{fo-ode}) and then separating variables, we obtain up to an additive integration constant,
\begin{equation}
\label{ab-sep}
	Z
	=
	\int\frac{v^{a-1}}{\sqrt{v^b-1}}{\rm d}v.
\end{equation}

Let us begin with the case in which $a=\kappa b$ for some integer $\kappa$. If $\kappa>0$ then
\[
	b(Z-Z_0)
	=
	\sqrt{v^b-1}
	\sum_{j=0}^{\kappa-1}
	\frac 2{2j+1}
		\left(
			\begin{matrix}
						\kappa-1 \\ j
					\end{matrix}
		\right)
		\left(v^b-1\right)^{j},
\]
where $Z_0$ is a constant.
So $v$ is an algebraic function of $Z$.
If $\kappa=0$, then
\[
	v^b=\left(\sec\left[\frac b2\left(Z-Z_0\right)\right]\right)^2.
\]
So $v$ is finitely many-valued. 

For $k=1,2,\ldots$, we have the recurrence relation
\[
	kb\int\frac{v^{-kb-1}}{\sqrt{v^b-1}}\,{\rm d} v
	=
	\frac{2k-1}2b\int\frac{v^{-(k-1)b-1}}{\sqrt{v^b-1}}\,{\rm d} v
	+
	v^{-kb}\sqrt{v^b-1}.
\]
It follows that if $\kappa=-\tilde\kappa<0$ then
\[
	Z-Z_0
	=\int\frac{v^{-\tilde\kappa b-1}}{\sqrt{v^b-1}}\,{\rm d} v
	=
	\sqrt{v^b-1}\,P_{\tilde\kappa}\left(v^{-b}\right)+c_{\tilde\kappa} \tan^{-1}\sqrt{v^b-1},
\]
where $c_{\tilde\kappa}\ne 0$ and $P_{\tilde\kappa}$ is a polynomial of degree $\tilde\kappa$.
It follows that in this case $v$ is not finitely many-valued and so equation (\ref{fo-ode}) does not have the algebro-Painlev\'e property.

Now assume that $a/b\ne\mathbb{Z}$.  In particular, this implies that $b\ge 2$.
The integrand in equation (\ref{ab-sep}) is defined on a two-sheeted Riemann surface with square-root branch points at $1,\omega, \omega^2,\ldots\omega^{b-1}$,
where $\omega={\rm e}^{2\pi i/b}$.   Evaluating the integral along any closed curve containing exactly two of these branch points in its interior will result in a (possibly zero) period of $v$ as a function of $Z$.
In particular, we have the following half periods:
\[
	h_k=\int_{\omega^{k}}^{\omega^{k+1}}\frac{v^{a-1}}{\sqrt{v^b-1}}{\rm d}v,
\]
where each path of integration is an arc on the circle $|v|=1$, traced in the positive direction.
Using the identity
\[
	\int_0^\pi (\sin\theta)^{\alpha-1}{\rm e}^{{\rm i}\beta \theta}{\rm d}\theta
	=
	\frac{\pi{\rm e}^{{\rm i}\pi \beta/2}}{\alpha 2^{\alpha-1}B\left([\alpha+\beta+1]/2,[\alpha-\beta+1]/2\right)},
\]
where $\mbox{Re}\,\alpha>0$ and
\[
	B(x,y)=\frac{\Gamma(x)\Gamma(y)}{\Gamma(x+y)}
\]
is the beta function, we have that
\begin{equation}
\label{hk-eval}
	h_k=2\omega^{ka}{\rm e}^{{\rm i}\pi a/b}\sin\left(\frac{\pi a}b\right),
\end{equation}
which is non-zero since $a/b\not\in\mathbb{Z}$.

Now assume that  $b<2a$ (and still $a/b\not\in\mathbb Z$).  Then as $v\to\infty$, equation (\ref{ab-sep}) shows that $Z\sim v^\eta$, where $\eta=(2a-b)/2>0$.
So there is a steepest ascent curve $\gamma(t)$ in the $v$-plane  such that $\gamma(t)\to \infty$ as $t\to 1$ from below and $\gamma(t)=\rho(t) h_1$, where $\rho$ is an increasing real positive function, $\rho(t)\to\infty$ as $t\to 1$ from below.  Using the same argument as earlier, we see that $v$ is an infinitely many-valued function of $Z$.  We reach the same conclusion if we assume that $a<0$ by considering a steepest ascent curve $\gamma(t)=\rho(t) h_1$, with $\rho(t)\to 0$ as $t\to 1^-$.
We have proved the following.

\begin{lemma}
\label{lemma-abound}
Let $a$ and $b$ be integers, $b>0$.  If $v$ given by equation (\ref{ab-sep}) is finitely many-valued, then either $a/b$ is a non-negative integer or
$0\le 2a\le b$.
\end{lemma}

Consider the case $b=2$.  If $a$ is even then $a/b$ is an integer and we have already seen that this integer must be non-negative. If $a$ is odd, then Lemma \ref{lemma-abound} shows that $a=1$, in which case equation (\ref{ab-sep}) integrates to give
\[
	v(Z)=\cosh(Z-Z_0),
\]
for some constant $Z_0$.

From now on we assume that $b\ge 3$.
Returning to equation (\ref{hk-eval}),
we observe that the half-periods satisfy
$h_1=\omega^ah_0$ and $h_{-1}=\bar\omega^a h_0$.
Therefore we have another half-period that is a real multiple of $h_0$, namely
\[
	\hat h:=h_1+h_{-1}=\sigma h_0,
	\quad\mbox{where}\quad\sigma=2\cos\left(\frac{2a\pi}{b}\right).
\]

Since $h_0$ and $\hat h$ are both half-periods, then $2mh_0+2n\hat h=2(m+n\sigma)h_0$ is a period.  Suppose now that $\sigma$ is irrational.  Then given any $x\in\mathbb{R}$ and any $\epsilon>0$, there are integers $m$ and $n$ such that $|x-m-n\sigma|<\epsilon$.  So given any $z_1\in\mathbb{C}$ and any $x_\infty\in \mathbb{R}$, there exists a sequence $(x_n)\subset\mathbb{R}$ such that $Z_n=z_1+x_nh_0\to Z_\infty:=z_1+x_\infty h_0$ and $v(Z_n)=v(Z_1)$ for all $n$.  
But this is not possible if the Riemann surface for $v$ as a function of $Z$ has only a finite number of sheets.  Therefore a necessary condition for equation (\ref{fo-ode}) to have the algebro-Painlev\'e property is that $\cos(2a\pi/b)$ be rational.  To find the choices of $a$ and $b$ such that this is true, we turn to Niven's Theorem, which is an elementary result from number theory.

\begin{theorem}
[Niven  \cite{niven:56}]\label{niven}
For $\phi\in [0,\pi/2]$, $\sin\phi$ and $\phi/\pi$ are both rational if and only if $\phi$ is $0$, $\pi/6$ or $\pi/2$.
\end{theorem}
Using the symmetries of cosine and the fact that $\cos\phi=\sin(\phi-\frac\pi 2)$, it follows that for $\phi\in\mathbb{R}$,  $\cos \phi$ and $\phi/\pi$ are both rational if and only if $|\cos\phi\,|$ is $0$, $1/2$ or $1$.
So a necessary condition for equation (\ref{fo-ode}) to have the algebro-Painlev\'e property is that $\cos(2a\pi/b)$ takes one of the values $\pm 1$, $\pm 1/2$ or $0$.  
Therefore Lemma \ref{lemma-abound} shows that if $a/b\not\in\mathbb{Z}$, then $2a/b=1/3,1/2,2/3$ or $1$.
We will now consider these cases in more detail.

If $2a/b=2/3$, then in terms of $w=v^{b/3}$ and $\hat z=bZ/6$, equation (\ref{ab-sep}) becomes
\begin{equation}
\label{w-gen}
	\hat z -\hat z_0=\frac12\int\frac{{\rm d}w}{\sqrt{w^3-1}}=\wp^{-1}(w;0,4),
\end{equation}
where $\hat z_0$ is a constant and $\wp(\zeta;g_2,g_3)$ is the Weiestrass elliptic function, which is satisfies $(\wp_\zeta)^2=4\wp^3-g2 \wp-g_3$, $\wp(0)=\infty$.
Similarly, if $2a/b=1/3$, then in terms of $w=v^{-b/3}$ and $\hat z={\rm i}bZ/6$, equation (\ref{ab-sep}) again becomes equation (\ref{w-gen}). Hence in both these cases, $w$ is a meromorphic function of $\hat z$.

If $2a/b=1/2$ then applying the transformation $w=v^{b/4}$ and $\hat z={\rm i} bZ/4$, $\varepsilon^2=-1$ to the integral (\ref{ab-sep}) gives
\[
\hat z-\hat z_0=\int\frac{{\rm d}w}{\sqrt{1-w^4}}=\mbox{sn}^{-1}(w;{\rm i}),
\]
where $\hat z_0$ is a constant and $\mbox{sn}(\zeta;k)$ is Jacobi elliptic function of elliptic modulus $k$, which satisfies
$(\mbox{sn}_\zeta)^2=(1-\mbox{sn}^2)(1-k^2\mbox{sn}^2)$, $\mbox{sn}(0;k)=0$.  Hence $w$ is a single-valued function of $\hat z$.

Finally if $2a/b=1$, then in terms of $w=v^{-b/3}$ and $\hat z={\rm i}bZ/6$, equation (\ref{ab-sep}) becomes
\[
	\hat z -\hat z_0=\frac12\int\frac{{\rm d}w}{w\sqrt{w^3-1}}=\frac23\tan^{-1}\sqrt{w^3-1},
\]
where $\hat z_0$ is a constant. So $w$ is a three-valued function of $\hat z$.

We have proved the following.

\begin{theorem}
\label{first-order-thm}
Let $s>0$ and $s$ be rational numbers.  Then the maximal analytic continuation of every solution of equation (\ref{fo-ode}) is finitely-valued
if and only if $(r+2)/s\in X$, where
\begin{equation}
\label{Xdef}
	X
	=
	\left\{0,\frac13,\frac12,\frac23,1\right\}
	\cup
	\{2p:p=1,2,3,\ldots\}.
\end{equation}
For element of $X$, the non-constant solutions are given below, where $z_0$ is an arbitrary constant.
\begin{itemize}
\item
When $(r+2)/s=0$ (i.e., when $r=-2$),
\[
u(z)=	\left\{
		\sec\left(\frac s2[z-z_0]\right)
	\right\}^{2/s}.
\]
\item
When $(r+2)/s=1/3$,
\[
u(z)=	\left\{
			\wp\left(\frac {{\rm i} s}6[z-z_0],0,4\right)
		\right\}^{-3/s}.
\]
\item
When $(r+2)/s=1/2$,
\[
u(z)=	\left\{
			\mbox{\rm sn}\left({\rm i}\frac s4[z-z_0],{\rm i}\right)
		\right\}^{4/s}.
\]
\item
When $(r+2)/s=2/3$,
\[
u(z)=	\left\{
			\wp\left(\frac s6[z-z_0],0,4\right)
		\right\}^{3/s}.
\]
\item
When $(r+2)/s=1$,
\[
u(z)
=
	\left\{
		\mbox{\rm cosh}\left(\frac s2[z-z_0]\right)
	\right\}^{2/s}.
\]
\item
When $(r+2)/s=2p$, $p=1,2,\ldots$, 
	\[
		\sqrt{u^s-1}\sum_{q=0}^{p-1}\frac 2{2q+1}
				\left(
					\begin{matrix}
						p-1 \\ q
					\end{matrix}
				\right)
				\left(u^s-1\right)^q=s(z-z_0).
	\]
\end{itemize}
\end{theorem}

\section{The Lane-Emden Equation}
\label{sec:laneemden}
There are three values of $n\ge 0$ for which the solution of the initial value problem (\ref{laneemden}--\ref{BCs}) is known to be explicitly solvable.  When $n=0$ or $n=1$, equation (\ref{laneemden}) is linear and the solutions are $\theta_0(\xi)=1-(\xi^2/6)$ and $\theta_1(\xi)=(\sin\xi)/\xi$ respectively.  The third solution, corresponding to $n=5$, is
$\theta_5(\xi)=1/\sqrt{1+(\xi^2/3)}$.  It is important for our analysis to observe that we can in fact find the general solution of equation (\ref{laneemden}) in these three cases.
The general solution $\theta_n$ of equation (\ref{laneemden}) in the cases $n=0$, $n=1$ and $n=5$ are respectively
\begin{equation*}
	\begin{split}
		\theta_0(\xi)&=-\frac{\xi^2}{6}+c_1+\frac{c_2}{\xi^2},
		\\
		\theta_1(\xi)&=(c_1\cos\xi+c_2\sin\xi)/\xi,
		\\
		\theta_5(\xi)&=\sqrt{\xi^{-1}w(z)},
	\end{split}
\end{equation*}
where $z=\ln\xi$,
$c_1$ and $c_2$ are integration constants
and $w$ satisfies
\begin{equation}
\label{w-elli[tic}
	\left(\frac{{\rm d}w}{{\rm d}z}\right)^2=\frac43w^4+w^2+c_1w.
\end{equation}
If $c_1\not\in\{0,{\rm i}/3,-{\rm i}/3\}$ then any non-constant solution $w$ of equation (\ref{w-elli[tic}) is an elliptic function. 
The remaining solutions are rational functions of exponentials.
It follows that the Lane-Emden equation (\ref{laneemden}) possesses the Painlev\'e property for $n=0$ and $n=1$ but not for $n=5$.  However, if we were to rewrite equation (\ref{laneemden}) for the new dependent variable $u(\xi):=\theta(\xi)^2$, then the resulting equation would have the Painlev\'e property.  In fact in this case the general solution has an infinitely-sheeted branch point at $\xi=0$, which is also an accumulation point of poles.

We will now analyse the singularity structure of solutions of the Lane-Emden equation (\ref{laneemden}) and show that for real $n\ge 0$, there are no other cases than those just described for which the equation possesses the algebro-Painlev\'e property.  We will use a natural generalisation of Painlev\'e's alpha test.
Choose any $\xi_0\ne 0$.  Then in terms of the non-zero parameter $\alpha$, we introduce the change of independent and dependent variable given by
\begin{equation}
\label{transform}
	\Theta(z)=\alpha^{2/(n-1)}\theta(\xi),\qquad z=(\xi-\xi_0)/\alpha.
\end{equation}
Under this transformation the Lane-Emden equation (\ref{laneemden}) becomes
\begin{equation}
\label{LE-alpha}
	\frac{{\rm d}^2\Theta}{{\rm d}z^2}+\frac{2\alpha}{\xi_0+\alpha z}\frac{{\rm d}\Theta}{{\rm d}z}+\Theta^n=0.
\end{equation}
Note that whenever $\alpha\ne 0$, equation (\ref{LE-alpha}) is equivalent to the Lane-Emden equation (\ref{laneemden}) and either both equations will have the algebro-Painlev\'e property or neither will.  We now consider solutions $\Theta_0(z)$ of equation (\ref{LE-alpha}) with $\alpha=0$, namely
\begin{equation}
\label{LE0}
	\frac{{\rm d}^2\Theta_0}{{\rm d}z^2}+\Theta_0^n=0.
\end{equation}
If the Lane-Emden equation (\ref{laneemden}) possesses the algebro-Painlev\'e property, then so does equation (\ref{LE0}). Observe that equation (\ref{LE0}) is autonomous, so it has no fixed singularities.  We will now show that apart from a finite number of values of $n$, equation (\ref{LE0}) has solutions that must live on infinitely-sheeted Riemann surfaces.

If $\Theta_0(z)$ is not a constant, then multiplying equation  (\ref{LE0}) through by $\Theta_0'(z)$ and integrating gives
\begin{equation}
\label{LE0int}
	\left(\frac{{\rm d}\Theta_0}{{\rm d}z}\right)^2=\frac{2}{n+1}\left(\kappa-\Theta_0^{n+1}\right),
\end{equation}
where $\kappa$ is an integration constant.
When $\kappa=0$, then general solution of equation (\ref{LE0int}) is
\[
	\Theta_0(z)=C(z-z_0)^{2/(1-n)},
\]
where $z_0$ is a constant and $C^{n-1}=2(n+1)(n-1)^{-2}$.
If $n$ is irrational then $\Theta_0$ has an infinitely-sheeted branch point at the movable singularity at $z=z_0$.  So now we assume that $n\ge 0$ is rational
and $\kappa\ne 0$.
Equation (\ref{LE0}) clearly has the algebro-Painlev\'e property for $n=0$ and $n=1$.  According to Theorem \ref{first-order-thm} with $r=0$ and $s=n+1$, the only other non-negative values of $n$ for which equation (\ref{LE0}) has the algebro-Painlev\'e property are $n=2$, $n=3$ and $n=5$.  
So in order to prove our claim, it only remains to show that equation (\ref{laneemden}) does not have the algebro-Painlev\'e property when $n=2$ and when $n=3$.  In fact, we will show in both these cases that the Lane-Emden equation (\ref{laneemden}) has solutions with movable logarithmic branch points, indicating that the solutions are locally, as well as globally, infinitely many-valued.

We will begin by considering the case $n=2$.  We look for a solution of the form
\begin{equation}
\label{n2ansatz}
	\theta(\xi)
	=
	\sum_{j=0}^\infty a_j(x)(\xi-\xi_0)^{j-2},\qquad x=\ln(\xi-\xi_0),
\end{equation}
where $\xi_0\in\mathbb{C}\setminus\{0\}$ and for all $j\in\{0,1,\ldots\}$, $a_j(x)$ is a polynomial in $x$.
Substituting the ansatz (\ref{n2ansatz}) in equation (\ref{laneemden}) with $n=2$ gives, at leading order, 
$\ddot a_0-5\dot a_0+6a_0+a_0^2=0$, whereas the coefficient of $(\xi-\xi_0)^{j-4}$ for $j>0$ gives
\[
		\xi_0
			\left[
				\ddot a_j+(2j-5)\dot a_j+(j-2)(j-3)a_j+\sum_{k=0}^j a_ka_{j-k}
			\right]
\]
\vskip -3mm
\[
		+
			\left[
				\ddot a_{j-1}+(2j-5)\dot a_{j-1}+(j-2)(j-3)a_{j-1}+\sum_{k=0}^{j-1} a_ka_{j-1-k}
			\right]=0,
\]
where dots denote differentiation with respect to $x=\ln(\xi-\xi_0)$.
Equating the coefficients of powers $(\xi-\xi_0)$ to zero, and recalling that each $a_j$ is a polynomial in $x$, we have
\[
	a_0(x)=-6,\quad  a_1(x)=\frac{12}{5}\xi_0^{-1},\quad  a_2(x)=-\frac{48}{25}\xi_0^{-2},
	\quad 
	a_3(x)=\frac{204}{125}\xi_0^{-3},
\]
\[
	a_4(x)=-\frac{876}{625}\xi_0^{-4},
	\quad
	a_5(x)=\frac{3612}{3125}\xi_0^{-5}
	\quad
	a_6(x)=c-\frac{1872}{21875}\xi_0^{-6}\log(\xi-\xi_0),
\]
where $c$ is an arbitrary constant.  The fact that $a_6$ is not a constant reflects the fact that the resonance condition of the existence of a locally meromorphic solution is not satisfied and that the equation does not pass the Painlev\'e test. For fixed $c$ and for all $j>6$, the polynomials $a_j$ are determined uniquely.
Hence the solution (\ref{n2ansatz}) has a logarithmic branch point at $\xi=\xi_0$, so it is locally, and therefore globally, infinitely branched.  Therefore the case $n=2$ does not have the algebro-Painlev\'e property.  A similar analysis shows that the case $n=3$ also has a logarithmic branch point.
This completes the proof of Theorem \ref{lane-emden-thm}.

\section{A general family of equations}
\label{sec:general}

In this section we will find necessary conditions for equations of the form (\ref{general-ode}) to possess the algebro-Painlev\'e property.  

\begin{theorem}
Let $\Omega$ be an open subset of $\mathbb{C}$ and let $L(z,w)$, $M(z,w)$ and $N(z,w)$ be rational functions of $w$ with coefficients that are analytic functions of $z$ in $\Omega$.
If equation (\ref{general-ode})
possesses the algebro-Painlev\'e property, then there is a non-empty open set $\tilde\Omega\subset\Omega$, rational numbers $l_1,\ldots,l_p$ and functions $a_1(z),\ldots,a_p(z)$ analytic on $\tilde\Omega$ such that
\begin{equation}
\label{Lform}
	L(z,w)=\sum_{i=1}^p\frac{l_i}{w-a_i(z)}.
\end{equation}
\end{theorem}

\noindent{\bf Proof.}
If $L(z,w)\equiv 0$ then we are done.  Now assume that $L(z,w)\not\equiv 0$ and choose $z_0\in\Omega$.
For $\alpha\ne 0$, let $Z=\alpha^{-1}(z-z_0)$ and $W(Z)=w(z)$.  Equation (\ref{general-ode}) now takes the form
\[
	\frac{{\rm d}^2W}{{\rm d}Z^2}=L(z_0+\alpha Z,W)\left(\frac{{\rm d}W}{{\rm d}Z}\right)^2+\alpha M(z_0+\alpha Z,W)\frac{{\rm d}W}{{\rm d}Z}+\alpha^2N(z_0+\alpha Z,W).
\]
So in the limit $\alpha\to 0$ we have
\[
	\frac{{\rm d}^2W}{{\rm d}Z^2}=L(z_0,W)\left(\frac{{\rm d}W}{{\rm d}Z}\right)^2,
\]
which can be written as the system
\begin{equation}
	\begin{split}
		\frac{{\rm d}W}{{\rm d}Z}
		&=
		P,
		\\
		\frac{{\rm d}P}{{\rm d}Z}
		&=
		L(z_0,W)P^2.
	\end{split}
	\label{WPsystem}
\end{equation}
For $W\to\infty$, $L(z_0,W)=\Lambda W^r\left(1+O\left(1/W\right)\right)$, for some integer $r$ and some non-zero constant $\Lambda$.
On making the substitution $\widetilde W(Z)=\alpha W(Z)$ and $\widetilde P(Z)=\alpha^{-r}P(Z)$, the system (\ref{WPsystem}) has the form
\begin{equation}
	\begin{split}
		\frac{{\rm d}\widetilde W}{{\rm d}Z}
		&=
		\alpha^{r+1}\widetilde P,
		\\
		\frac{{\rm d}\widetilde P}{{\rm d}Z}
		&=
		\Lambda \widetilde W^r\left(1+O\left(\alpha\right)\right)\widetilde P^2.
	\end{split}
	\label{WPtildesystem}
\end{equation}
Assume that $r\ge 0$. Solving  this system subject to the initial condition $\widetilde W(Z_0)=\widetilde W_0$ and $\widetilde P(Z_0)=\widetilde P_0$, where $\widetilde W_0$ and $\widetilde P_0$ are independent of $\alpha$, gives
\begin{equation*}
	\begin{split}
	W(Z)
	&=
	\widetilde W_0-\frac{\alpha^{r+1}}{\Lambda\widetilde W_0^{r}}\log
						\left(
								1-\Lambda\widetilde W_0^{r}\widetilde P_0(Z-Z_0)
						\right)+O\left(\alpha^{r+2}\right),
	\\
	P(Z)
	&=
	\frac{\widetilde P_0}{1-\Lambda\widetilde W_0^{r}\widetilde P_0(Z-Z_0)}+O(\alpha),
	\end{split}
\end{equation*}
which has a logarithmic branch point whenever $\widetilde W_0$ and $\widetilde P_0$ are non-zero. So the algebro-Painlev\'e property demands that $r\le-1$.  In other words, as a functions of $W$, the degree of the denominator of $L(z_0,W)$ is greater than the degree of the numerator.

Let us next consider a pole at $W=a$ of $L(z_0,W)$ of multiplicity $r>0$.  Near $W=a$, the system (\ref{WPsystem}) now has the form
\begin{equation}
	\begin{split}
		\frac{{\rm d} W}{{\rm d}Z}
		&=
		P,
		\\
		\frac{{\rm d} P}{{\rm d}Z}
		&=
		\Lambda  (W-a)^{-r}\left(1+O\left(W-a\right)\right) P^2,
	\end{split}
	\label{newWPsystem}
\end{equation}
for some $\Lambda\ne 0$.
For $\alpha\ne 0$ we let $\widetilde W(Z)=\alpha^{-1} (W(Z)-a)$ and $\widetilde P(Z)=\alpha^{-r}P(Z)$.
Then the system (\ref{newWPsystem}) becomes
\begin{equation}
	\begin{split}
		\frac{{\rm d}\widetilde W}{{\rm d}Z}
		&=
		\alpha^{r-1}\widetilde P,
		\\
		\frac{{\rm d}\widetilde P}{{\rm d}Z}
		&=
		\Lambda \widetilde W^{-r}\left(1+O(\alpha)\right)\widetilde P^2.
	\end{split}
	\label{newWPtildesystem}
\end{equation}
If $r>1$, we find the solution to the initial value problem (\ref{newWPtildesystem}) with $\widetilde W(Z_0)=\widetilde W_0$ and $\widetilde P(Z_0)=\widetilde P_0$, where $\widetilde W_0$ and $\widetilde P_0$ are independent of $\alpha$, is
\begin{equation*}
	\begin{split}
	W(Z)
	&=
	\widetilde W_0-\alpha^{r-1}\Lambda\widetilde W_0^{-r}\log
						\left(
								\frac{\Lambda-\widetilde W_0^{r}\widetilde P_0(Z-Z_0)}{\Lambda}
						\right)+O\left(\alpha^{r}\right),
	\\
	P(Z)
	&=
	\frac{\Lambda\widetilde P_0}{\Lambda-\widetilde W_0^{r}\widetilde P_0(Z-Z_0)}+O(\alpha),
	\end{split}
\end{equation*}
which has a logarithmic branch point if $\widetilde W_0$ and $\widetilde P_0$ are non-zero.  It follows that $r=1$, so the only poles that $L(z_0,W)$ can have are simple.  
Eliminating $\widetilde P$ from the system (\ref{newWPtildesystem}) with $r=1$ gives
\[
	\frac{\widetilde W''}{\widetilde W'}=\Lambda \frac{\widetilde W'}{\widetilde W}+O(\alpha).
\]
So either $\Lambda=1$ or
\[
	\widetilde W(Z)=\kappa(Z-C)^{1/(1-\Lambda)},
\]
where $\kappa$ and $C$ are constants.
It follows that $\widetilde W(Z)$ is infinitely branched around $Z=C$ unless $\Lambda$ is rational.
Combined with the fact that $L(z_0,W)$ has a zero at infinity, we conclude that it has the form
\[
	L(z_0,w)=\sum_{i=1}^p\frac{\tilde l_i}{w-\tilde a_i},
\]
for some constants $\tilde l_1,\ldots,\tilde l_p$ and $\tilde a_1,\ldots,\tilde a_p$.
Therefore $L(z,w)$ has the form
\[
	L(z,w)=\sum_{i=1}^p\frac{l_i(z)}{w-a_i(z)},
\]
where $l_i(z)$ and $a_i(z)$, $i=1,\ldots,p$, are analytic functions of $z\in\tilde\Omega$.  However, the $l_i(z)\in\mathbb{Q}$ for all $i\in\{1,\ldots,p\}$ and all $z\in\tilde \Omega$, so therefore they are constant rational numbers.  
\hfill$\Box$

\vskip 3mm

Next we consider the case in which $M(z,w)$ and $N(z,w)$ do not both vanish.
By making a transformation of the form $w(z)\mapsto w(z)+a(z)$, if necessary, we can assume that a pole of $L$, $M$ or $N$ is at $w=0$.

\begin{theorem}
\label{thm-geneqn-zero}
Let $\Omega$ be an open subset of $\mathbb{C}$ and let $L(z,w)$, $M(z,w)$ and $N(z,w)$ be rational functions of $w$ with coefficients that are analytic functions of $z$ in $\Omega$.
Assume that $M(z,w)$ and $N(z,w)$ are not both identically zero.
Suppose that there exists $l\in{\mathbb{Q}}$, $m\in\mathbb{Z}$,  $n\in\mathbb{Z}$ and analytic functions $M_0(z)$ and $N_0(z)$ on $\Omega$ such that for all $z\in\Omega$,
\[
	L(z,w)=\frac lw+O(1),
	\ 
	M(z,w)=\frac{M_0(z)}{w^m}\left[1+O(w)\right],
\]
\[
	\mbox{and\ \ }
	N(z,w)=\frac{N_0(z)}{w^n}\left[1+O(w)\right],
	\mbox{as }w\to 0.
\]

Assume that equation (\ref{general-ode}) possesses the algebro-Painlev\'e property.

\noindent{\it (i)}
If $n\ge 0$, $N_0(z)\ne 0$ for all $z\in \Omega$ and either $M(z,w)\equiv 0$ or $n>2m-1$, then
$n+2l=1$ or
\[
	n+2l>1
	\mbox{\ \ and\ \ }
	\frac{n+1}{n+2l-1}\in X
\]
or
\[
	n+2l<1
	\mbox{\ \ and\ \ }\frac{2(l-1)}{n+2l-1}\in X,
\]
where $X=\left\{0,\frac13,\frac12,\frac23,1\right\}
	\cup
	\{2p:p=1,2,3,\ldots\}$.

\noindent{\it (ii)}
If $m\ge 1$, $M_0(z)\ne 0$ for all $z\in \Omega$ and either $N(z,w)\equiv 0$ or $n<2m-1$, then
$m+l=1$ or
\[
	m+l>1
	\mbox{\ \ and\ \ }
	\frac{m}{l+m-1}\in\{0,1/2,1\}
\]
or
\[
	m+l<1
	\mbox{\ \ and\ \ }\frac{l-1}{l+m-1}\in\{0,1/2,1\}.
\]

\end{theorem}

\noindent{\bf Proof.}
{\it (i)} For $\alpha\ne 0$ and $z_0\in \Omega$, let $Z:=(z-z_0)/\alpha^{n+1}$
and $W(Z):=\alpha^{-2}w(z)$ in equation (\ref{general-ode}), giving
\[
	\frac{{\rm d}^2W}{{\rm d}Z^2}
	=\left(\frac{l}{W}+O(\alpha)\right)\left(\frac{{\rm d}W}{{\rm d}Z}\right)^2
\]
\[
	+\alpha^{n+1-2m}\frac{M_0(z_0)}{W^m}\left(1+O(\alpha)\right)\frac{{\rm d}W}{{\rm d}Z}
	+\frac{N_0(z_0)}{W^n}\left(1+O(\alpha)\right).
\]
Since $M_0=0$ or $n> 2m-1$ then in the limit $\alpha\to 0$, this equation becomes
\begin{equation}
\label{n-dom}
	\frac{{\rm d}^2W}{{\rm d}Z^2}
	=\frac{l}{W}\left(\frac{{\rm d}W}{{\rm d}Z}\right)^2
	+\frac{N_0(z_0)}{W^n}.
\end{equation}
With the help of the integrating factor $W^{-l}$, we see that, if $n+2l\ne 1$, then any non-constant solution satisfies
\begin{equation}
\label{n-dom-gen}
	(W')^2=\frac{N_0(z_0)}{n+2l-1}\frac{CW^{n+2l-1}-1}{W^{n-1}},
\end{equation}
where $C$ is a constant.

When $C\ne 0$, equation (\ref{n-dom-gen}) can be rescaled to take the form of equation (\ref{fo-ode}) with $r=n-1$ and $s=n+2l-1$.  From Theorem \ref{first-order-thm}, we see that when
$s=n+2l-1>0$, $(r+2)/s=(n+1)/(n+2l-1)\in X$. When $s=n+2l-1<0$, we define $\tilde r=r-s=-2l$ and $\tilde s=-s=1-n-2l$, and the required condition is $(\tilde r+2)/\tilde s=2(l-1)/(n+2l-1)\in X$.

\vskip 5 mm

{\it (ii)} 
For $\alpha\ne 0$ and $z_0\in  \Omega$, let $Z:=(z-z_0)/\alpha^{2m}$ and
$W(Z):=\alpha^{-2}w(z)$ in equation (\ref{general-ode}), giving
\[
	\frac{{\rm d}^2W}{{\rm d}Z^2}
	=\left(\frac{l}{W}+O(\alpha)\right)\left(\frac{{\rm d}W}{{\rm d}Z}\right)^2
	+\frac{M_0(z_0)}{W^m}\left(1+O(\alpha)\right)\frac{{\rm d}W}{{\rm d}Z}
\]
\[
	+\alpha^{2(2m-1-n)}\frac{N_0(z_0)}{W^n}\left(1+O(\alpha)\right).
\]
Since either $N_0=0$ or $n<2m-1$, then
in the limit $\alpha\to 0$, this becomes
\[
	\frac{{\rm d}^2W}{{\rm d}Z^2}
	=\frac{l}{W}\left(\frac{{\rm d}W}{{\rm d}Z}\right)^2
	+\frac{M_0(z_0)}{W^m}\frac{{\rm d}W}{{\rm d}Z}.
\]
Using the integrating factor $W^{-l}$, we find that when $l+m\ne 1$, then
\begin{equation}
\label{m-dom-gen}
	W'
	=
	\frac{M_0}{l+m-1}\frac{CW^{l+m-1}-1}{W^{m-1}},
\end{equation}
where $C$ is a constant.  When $C\ne 0$, equation (\ref{m-dom-gen}) can be rescaled to take the form of equation (\ref{fod-ode}) with $r=m-1$ and $s=l+m-1$. 
So from Corollary \ref{deg1-cor}, the algebro-Painlev\'e property requires that for $s=l+m-1>0$, we must have $(r+1)/s=m/(l+m-1)\in\{1,1/2,1\}$.
When $s=l+m-1<0$, we let $\tilde r=r-s=-l$ and $\tilde s=-s=1-l-m$ and we must have 
$(\tilde r+1)/\tilde s=(l-1)/(l+m-1)\in\{1,1/2,1\}$.
\hfill$\Box$

\vskip 3mm

The following theorem concerns the asymptotic forms of $L$, $M$ and $N$ as $w\to\infty$.

\begin{theorem}
\label{thm-geneqn-pole}
Let $\Omega$ be an open subset of $\mathbb{C}$ and let $L(z,w)$, $M(z,w)$ and $N(z,w)$ be rational functions of $w$ with coefficients that are analytic functions of $z$ in $\Omega$.
Suppose that there exists $\hat l\in{\mathbb{Q}}$ such that for all $z\in\Omega$, as $w\to \infty$,
\[
	L(z,w)=\frac {\hat l}w+O\left(\frac1{w^2}\right),
	\ 
	M(z,w)=\frac{\hat M_0(z)}{w^m}\left[1+O\left(\frac1w\right)\right],
\]
\[
	N(z,w)=\frac{\hat N_0(z)}{w^n}\left[1+O\left(\frac1w\right)\right].
\]
Assume that equation (\ref{general-ode}) possesses the algebro-Painlev\'e property.

\noindent{\it (i)}
If $n\ge 0$, $\hat N_0(z)\ne 0$ for all $z\in \Omega$ and either $M(z,w)\equiv 0$ or $n>2m-1$, then
 $n+2l=1$ or
\[
	n+2l>1
	\mbox{\ \ and\ \ }
	\frac{n+1}{n+2l-1}\in X
\]
or
\[
	n+2l<1
	\mbox{\ \ and\ \ }\frac{2(l-1)}{n+2l-1}\in X,
\]
where $X=\left\{0,\frac13,\frac12,\frac23,1\right\}
	\cup
	\{2p:p=1,2,3,\ldots\}$.

\noindent{\it (ii)}
If $m\ge 1$, $\hat M_0(z)\ne 0$ for all $z\in \Omega$ and either $N(z,w)\equiv 0$ or $n<2m-1$, then
$m+l=1$ or
\[
	m+l>1
	\mbox{\ \ and\ \ }
	\frac{m}{l+m-1}\in\{0,1/2,1\}
\]
or
\[
	m+l<1
	\mbox{\ \ and\ \ }\frac{l-1}{l+m-1}\in\{0,1/2,1\}.
\]
\end{theorem}

\vskip 3 mm
\noindent{\bf Proof.}

\noindent{\it (i)}
For $\alpha\ne 0$ and $z_0\in \Omega$, let $Z:=(z-z_0)/\alpha^{n-1}$ and
$W(Z):=\alpha^2 w(z)$ in equation (\ref{general-ode}), giving
\[
	\frac{{\rm d}^2W}{{\rm d}Z^2}
	=\left(\frac{\hat l}{W}+O(\alpha)\right)\left(\frac{{\rm d}W}{{\rm d}Z}\right)^2
	+\frac{\hat M_0(z_0)}{W^m}\left(1+O(\alpha)\right)\frac{{\rm d}W}{{\rm d}Z}
\]
\[
	+\alpha^{2m+1-n}\frac{\hat N_0(z_0)}{W^n}\left(1+O(\alpha)\right).
\]
Since either $\hat N_0(z_0)=0$ or $n>2m+1$, then in the limit $\alpha\to 0$, we have
\[
	\frac{{\rm d}^2W}{{\rm d}Z^2}
	=\frac{\hat l}{W}\left(\frac{{\rm d}W}{{\rm d}Z}\right)^2
	+\frac{\hat N_0(z_0)}{W^n}.
\]
When $n+2l\ne 1$, we have
\[
	\left(\frac{{\rm d}W}{{\rm d}Z}\right)^2
	=
	\frac{\hat N_0(z_0)}{n+2l-1}\frac{Cw^{n+2l-1}-1}{w^{n-1}}.
\]
The result follows from Theorem \ref{first-order-thm}.

\noindent{\it (ii)}
For $\alpha\ne 0$ and $z_0\in \Omega$, let $Z:=(z-z_0)/\alpha^{m}$ and
$W(Z):=\alpha w(z)$ in equation (\ref{general-ode}), giving
\[
	\frac{{\rm d}^2W}{{\rm d}Z^2}
	=\left(\frac{\hat l}{W}+O(\alpha)\right)\left(\frac{{\rm d}W}{{\rm d}Z}\right)^2
	+\alpha^{n-2m-1}\frac{\hat M_0(z_0)}{W^m}\left(1+O(\alpha)\right)\frac{{\rm d}W}{{\rm d}Z}
\]
\[
	+\frac{\hat N_0(z_0)}{W^n}\left(1+O(\alpha)\right).
\]
Since either $\hat M_0(z_0)=0$ or $n<2m+1$, then in the limit $\alpha\to 0$, we have
\[
	\frac{{\rm d}^2W}{{\rm d}Z^2}
	=\frac{\hat l}{W}\left(\frac{{\rm d}W}{{\rm d}Z}\right)^2
	+\frac{\hat M_0(z_0)}{W^m}\frac{{\rm d}W}{{\rm d}Z}.
\]
Using the integrating factor $W^{-\hat l}$, we see that if $m+l\ne 1$,
\[
	\frac{{\rm d}W}{{\rm d}Z}
	=
	\frac{\hat M_0(z_0)}{m+l-1}\frac{CW^{m+l-1}-1}{W^{m-1}},
\]
where $C$ is a constant.
The result follows from Corollary \ref{deg1-cor}.
\hfill$\Box$

\vskip 5 mm

We finish by considering the following important case.
\begin{theorem}
\label{laurent-thm}
Consider the equation
\begin{equation}
\label{laurent}
	w''(z)=\sum_{j=\mu}^\nu a_j(z)w(z)^j,
\end{equation} 
where $\mu,\nu\in\mathbb{Z}$, $\mu\le\nu$ and there is a domain $\Omega$ on which $a_\mu,\ldots,a_\nu$ are analytic and $a_\mu(z)a_\nu(z)\ne 0$ for all $z\in\Omega$.
If equation (\ref{laurent}) has the algebro-Painlev\'e property then one of the following is true.
\begin{enumerate}
\item
If $\nu\le1$, then equation (\ref{laurent}) has the form
\begin{equation}
\label{almost-linear}
	w''(z)=c_0 \left\{w(z)^{-3}+\rho(z)w(z)^{-2}\right\}+\sigma(z)+\tau(z)w(z),
\end{equation}
where $c_0$ is a constant and $\rho$, $\sigma$ and $\tau$ are analytic on $\Omega$.
\item
If $\nu> 1$, then $\nu\in\{2,3,5\}$.  Under the change of variables
\begin{equation}
\label{CoV}
	w(z):=\phi(z)\tilde w(\tilde z)-\psi(z),
	\qquad
	\tilde z:=\int_{z_0}^z\frac{{\rm d}\zeta}{\phi(\zeta)^2},
\end{equation}
where
\begin{equation*}
\begin{split}
	\phi(z)^{\nu+3}&=\frac{2(\nu+1)}{(\nu-1)^2a_\nu(z)}\quad\mbox{and}
\\
	\psi(z)
		&=
		\begin{cases}
		\displaystyle
			\frac{a_1(z)-\phi(z)^{-1}\phi''(z)}{2a_2(z)},&\nu=2,
			\\
			\displaystyle 
			\frac{a_{\nu-1}(z)^{\ }}{\nu a_\nu(z)},& \nu=3\mbox{\ or\ }5,
		\end{cases}
\end{split}
\end{equation*}
equation (\ref{laurent}) becomes
\begin{equation}
\label{almost-painleve}
\begin{split}
	\tilde w_{\tilde z\tilde z}&=c_0\left\{\frac{1}{(\tilde w-\sigma(\tilde z))^3}+\frac{\rho(\tilde z)}{(\tilde w-\sigma(\tilde z))^2}\right\}
	+c_1+c_2\tilde z+6\tilde w^2,
	\\
	\tilde w_{\tilde z\tilde z}&=c_0\left\{\frac{1}{(\tilde w-\sigma(\tilde z))^3}+\frac{\rho(\tilde z)}{(\tilde w-\sigma(\tilde z))^2}\right\}
	+c_1+(c_2\tilde z+c_3)\tilde w+2\tilde w^3,\mbox{\ or}
	\\
	\tilde w_{\tilde z\tilde z}&=c_0\left\{\frac{1}{(\tilde w-\sigma(\tilde z))^3}+\frac{\rho(\tilde z)}{(\tilde w-\sigma(\tilde z))^2}\right\}
	\\
	&
	\ +\tau(\tilde z)+\left\{(c_1 +c_2 \tilde z)^2+c_3\right\}\tilde w+\chi(\tilde z)\tilde w^2+2(c_1+c_2\tilde z)\tilde w^3+\frac34\tilde w^5,
\end{split}
\end{equation}
where $c_0$, $c_1$, $c_2$ and $c_3$ are constants and $\rho$, $\sigma$, $\tau$ and $\chi$ are analytic on $\Omega$.
\end{enumerate}
\end{theorem}

Observe that when $c_0=0$, equations (\ref{almost-linear}), (\ref{almost-painleve}\,a) and (\ref{almost-painleve}\,b) have the Painlev\'e property, and hence the algebro-Painlev\'e property.  More precisely, when $c_0=0$, equation (\ref{almost-linear}) is linear, and so has no movable singularities, while equations (\ref{almost-painleve}\,a) and (\ref{almost-painleve}\,b) can be transformed to the first and second Painlev\'e equations,
\[
	u''(x)=6u(x)^2+x,\qquad u''(x)=2u(x)^3+xu(x)+\alpha,
\]
respectively, for some constant $\alpha$, provided that $c_2\ne 0$.  When $c_0=0=c_2$, equations (\ref{almost-painleve}\,a) and (\ref{almost-painleve}\,b) can be solved in terms of elliptic functions.  The other equations in Theorem \ref{laurent-thm} are not expected to possess the algebro-Painlev\'e property for arbitrary $c_0$, $\rho$, $\sigma$ and $\tau$.  However, there are choices of these coefficients that provide further non-trivial examples of equations with this property.  For example, introducing the change of variables $w(z)=u(x)^{1/2}$, $z=x$, in the fourth Painlev\'e equation
\[
	u''(x)=\frac12\frac{u'(x)^2}{u(x)}+\frac 32u(x)^3+4xu(x)^2+2(x^2-\alpha)u(x)+\frac{\beta}{u(x)},
\]
where $\alpha$ and $\beta$ are constants, results in the equation
\begin{equation}
\label{dnls-red}
	w''(z)=\frac{\beta}{2w(z)^3}+(z^2-\alpha)w(z)+2zw(z)^3+\frac34w(z)^5,
\end{equation}
which is a special case of equation (\ref{almost-painleve}\,c) that possesses the algebro-Painlev\'e property.  Equation (\ref{dnls-red}) is known to arise as a symmetry reduction of the derivative nonlinear Schr\"odinger equation (DNLS) \cite{ablowitzrs:80}
\[
	{\rm i}q_t+q_{xx}\pm{\rm i}\left(|q|^2q\right)_x=0.
\]
Connection formulas for equation (\ref{dnls-red}) have been derived and studied in \cite{bassomchm:92}, as well as B\"acklund transformations and exact solutions.

Several results in the literature go some way to addressing the question of which cases of equation (\ref{almost-painleve}\,c) possess the algebro-Painlev\'e property.  In the following result, the notation $T(r,f)$ denotes the Nevanlinna characteristic of the meromorphic function $f$ (see, e.g. Hayman \cite{hayman:64}).
\begin{theorem}
[Halburd and Kecker \cite{halburdk:14}]
\label{HK-thm}
Let $y$ be a solution of the equation 
$\displaystyle y'' = y^5 + \sum_{k=0}^4 a_k(z) y^k$,
where $y$ also satisfies 
$y(z)^2 + s_1(z) y(z) + s_2(z) = 0$,
where $s_1$, $s_2$, $a_0,\ldots,a_4$ are meromorphic functions satisfying
$T(r,a_j)=o\left(T(r,s_k)\right)$ as $r\to\infty$ outside some possible exceptional set of finite Lebesgue measure, for all $j\in\{1,2\}$ and $k\in\{1,2,3,4\}$.
Then $s_1$ can be expressed linearly in terms of $a_4$ and $s_2$ can be expressed either in terms of the fourth Painlev\'e transcendents or it solves a Riccati equation with coefficients that are polynomials in the $a_j$s and their derivatives.
\end{theorem}

The Hamiltonian system
\begin{equation}
\label{takasaki}
	\frac{{\rm d}q}{{\rm d}t} = p,
	\quad
	\frac{{\rm d}p}{{\rm d}t} = \frac{8\beta}{q^3}+\left(t^2-\alpha\right)q+\frac t2q^3+\frac 3{64}q^5,
\end{equation}
is equivalent to equation (\ref{dnls-red}) and was obtained in \cite{takasaki:01} as a second-order rational Painlev\'e-Calogero system.
In \cite{filipuks:23}, Filipuk and Stokes constructed a rational surface $\chi^{\mbox{Tak}}$ by first compactifying the the space of variable $(q,p)$ to $\mathbb{P}^1\times\mathbb{P}^1$ and then peforming a sequence of twenty blowups to desingularise the Takasaki system (\ref{takasaki}).  They also proved the following.
\begin{theorem}
\label{FS-thm}
	[Filipuk and Stokes \cite{filipuks:23}]
	If the system
\[
	q'=p,\qquad p'=\frac{3q^5}{64}+\sum_{k=-3}^4a_k(t)q^k,
\]
is quadratically regularisable on a bundle of rational surfaces obtained by blowing up points in the same configuration as $\chi^{\mbox{Tak}}$, then it must coincide with the Takasaki system up to affine changes of independent variable.
\end{theorem}
The notion of {\em quadratic regularisability} involves using a map $u\mapsto u^2$ in local coordinates in the blowing-up procedure in order to obtain a regular system.
From the point of view of our present work, one drawback of both Theorem \ref{HK-thm} and Theorem \ref{FS-thm} is that they only consider solutions that are globally double-sheeted. As the example of equation (\ref{abel}) illustrates, solutions of ODEs with globally finitely many-valued solutions with only square-root-type branch points may require a Riemann surface with more than two sheets.

\vskip 3 mm

\noindent{\bf Proof of Theorem \ref{laurent-thm}.}
From Theorem \ref{thm-geneqn-zero}, we see that either $\mu\ge 0$ or $\mu=-3$.  From Theorem \ref{thm-geneqn-pole} we see that either $\nu\le 3$ or $\nu=5$.
Therefore if $\nu\le 1$ then equation (\ref{laurent}) has the form
\begin{equation}
\label{Plinear}
	w_{zz}=P\left(z,1/w\right)+a_0(z)+a_1(z)w,
\end{equation}
where $P$ is either identically zero or a polynomial of degree exactly three in its second argument with coefficients that are analytic in $\tilde z$ in a neighbourhood $\widetilde\Omega$ satisfying $P(\tilde z,0)=0$.
Next we consider the case $\nu\ge 2$. It follows from Theorem \ref{thm-geneqn-pole} that $\nu\in\{2,3,5\}$.  
Under the transformation (\ref{CoV}), equation (\ref{laurent}) becomes
\begin{equation}
\label{pole-norm}
\begin{split}
	\tilde w_{\tilde z\tilde z}
	&=
	P\left(\tilde z,(\tilde w-\sigma(\tilde z))^{-1}\right)
	+
	\tilde a_0(\tilde z)+6\tilde w^2,
	\\
	\tilde w_{\tilde z\tilde z}
	&=
	P\left(\tilde z,(\tilde w-\sigma(\tilde z))^{-1}\right)
	+
	\tilde a_0(\tilde z)+\tilde a_1(\tilde z) \tilde w+2\tilde w^3,\mbox{\ \ \ or}
	\\
	\tilde w_{\tilde z\tilde z}
	&=
	P\left(\tilde z,(\tilde w-\sigma(\tilde z))^{-1}\right)
	+
	\tilde a_0(\tilde z)+\tilde a_1(\tilde z)\tilde w+\tilde a_2(\tilde z)\tilde w^2+\tilde a_3(\tilde z)\tilde w^3+\frac34\tilde w^5,
\end{split}
\end{equation}
where $\sigma(\tilde z)=\psi(z)/\phi(z)$ and again $P$ is a polynomial as described above. 
If $P$ is not identically zero, we write $P(\tilde z,W)=a_{-3}(\tilde z)W^3+a_{-2}(\tilde z)W^2+a_{-1}(\tilde z)W$.
We look for solutions of equations (\ref{Plinear}) and (\ref{new-pole-norm}) in a neighbourhood of a point $\tilde z_0$ such that $\tilde w\sim c_0(\tilde z-\tilde z_0)^q$, for some $q>0$. (For equation (\ref{Plinear}) we take $\tilde z=z$, $\tilde w(\tilde z)=w(z)$ and $\sigma(\tilde z)=0$.)
Equations (\ref{Plinear}) and (\ref{new-pole-norm}) admit solutions of the form
\[
	\tilde w(\tilde z)=\sum_{n=0}^\infty u_n(x)\zeta^{(n+1)/2},
	\qquad
	\zeta=\tilde z-\tilde z_0,
\]
where $u_n$ is a polynomial in $x=\log\zeta$. At leading order we have $u_0^4=-4a_{-3}(z_0)$. The two choices for $u_0^2=\pm2\sqrt{-a_{-3}(z_0)}$ correspond to solutions with different leading-order terms.  The choices $\pm u_0$ correspond to different branches of the leading order term $u_0\zeta^{1/2}$ of the same solution.
The first few terms in the series expansion are

\[
	\tilde w(\tilde z)-\sigma(\tilde z)
	=
	c_0\zeta^{1/2}+\frac{c_0^2}{3}\frac{\tilde a_{-2}(\tilde z_0)}{\tilde a_{-3}(\tilde z_0)}\zeta
	+\frac 2{c_0}	\left\{
					\kappa+A(z_0)\log\zeta
				\right\}\zeta^{3/2}+\cdots,
\]
where $A(z_0)=\tilde a'_{-3}(\tilde z_0)u_0^2-4\tilde a_{-1}(\tilde z_0)\tilde a_{-3}(\tilde z_0)$ and $\kappa$ is a constant.
In order to avoid a solution requiring a Riemann surface with an infinite number of sheets, we require that
\[
	\tilde a'_{-3}(\tilde z_0)u_0^2-4\tilde a_{-1}(\tilde z_0)\tilde a_{-3}(\tilde z_0)=0.
\]
This condition must hold for all $\tilde z_0\in\tilde\Omega$ and for both choice of $u_0^2=\pm2\sqrt{-a_{-3}(z_0)}$.
Recall that $\tilde a_{-3}\ne 0$, so  $\tilde a_{-1}(\tilde z)=0$ and $\tilde a_{-3}(\tilde z)=c_0$, for some non-zero constant $c_0$.
Writing $\tilde a_{-2}(\tilde z)=c_0\rho(\tilde\zeta)$, then equations (\ref{pole-norm}) become
\begin{equation}
\label{new-pole-norm}
\begin{split}
	\tilde w_{\tilde z\tilde z}
	&=
	c_0\left\{\frac{1}{(\tilde w-\sigma(\tilde z))^3}+\frac{\rho(\tilde z)}{(\tilde w-\sigma(\tilde z))^2}\right\}
	+
	\tilde a_0(\tilde z)+6\tilde w^2,
	\\
	\tilde w_{\tilde z\tilde z}
	&=
	c_0\left\{\frac{1}{(\tilde w-\sigma(\tilde z))^3}+\frac{\rho(\tilde z)}{(\tilde w-\sigma(\tilde z))^2}\right\}
	+
	\tilde a_0(\tilde z)+\tilde a_1(\tilde z) \tilde w+2\tilde w^3,\mbox{\ \ \ or}
	\\
	\tilde w_{\tilde z\tilde z}
	&=
	c_0\left\{\frac{1}{(\tilde w-\sigma(\tilde z))^3}+\frac{\rho(\tilde z)}{(\tilde w-\sigma(\tilde z))^2}\right\}
	\\
	&\qquad
	+
	\tilde a_0(\tilde z)+\tilde a_1(\tilde z)\tilde w+\tilde a_2(\tilde z)\tilde w^2+\tilde a_3(\tilde z)\tilde w^3+\frac34\tilde w^5,
\end{split}
\end{equation}
where, allowing $c_0$ to be zero, covers both the case when $P$ is of degree exactly three and when $P$ is identically zero.
Also, equation (\ref{Plinear}) becomes equation (\ref{almost-linear}).

Next we study equations (\ref{new-pole-norm}) in the neighbourhood of a point $\tilde z_0$ where $\tilde w\sim u_0(\tilde z-\tilde z_0)^{-q}$, for $q>0$.
Equation (\ref{new-pole-norm}\,a) admits a solution of the form
\[
	\tilde w(\tilde z)=\sum_{n=0}^\infty u_n(x)\zeta^{n-2},
	\qquad
	\zeta=\tilde z-\tilde z_0,
\]
where $u_n$ is a polynomial in $x=\log\zeta$ satisfying a recurrence relation of the form
\[
	(n+1)(n-6)u_n+(2n-5)\dot u_n+\ddot u_n=F_n(u_0,\ldots,u_{n-1}),
	\qquad
	u_0=1,
\]
where $F_n$ is a polynomial in its arguments.  The first few terms in the series expansion are
\[
	\tilde w(\tilde z)
	=
	\frac1{\zeta^2}-\frac{\tilde a_0(\tilde z_0)}{10}\zeta^2-\frac{\tilde a_0'(\tilde z_0)}6\zeta^3+
	\left\{\kappa-\frac{\tilde a_0''(\tilde z_0)}{14}\log\zeta\right\} \zeta^6+\cdots,
\]
where $\kappa$ is a constant.  The presence of the logarithmic term indicates that the solution is infinitely-many valued in a neighbourhood of $\tilde z_0$.  Hence a necessary condition for the equation to posses the algebro-Painlev\'e property is that $\tilde a_0''(\tilde z_0)=0$ for all $\tilde z_0\in\Omega$. Therefore $\tilde a_0(\tilde z)=c_1+c_2\tilde z$, for some constants $c_1$ and $c_2$. Hence equation (\ref{new-pole-norm}\,a) becomes equation (\ref{almost-painleve}\,a).

Similarly, equation (\ref{new-pole-norm}\,b) admits a solution of the form
\[
	\tilde w(\tilde z)=\sum_{n=0}^\infty u_n(x)\zeta^{n-1},
\]
where $u_n$ is a polynomial in $x=\log\zeta$ satisfying a recurrence relation of the form
\[
	(n+1)(n-4)u_n+(2n-3)\dot u_n+\ddot u_n=G_n(u_0,\ldots,u_{n-1}),
	\qquad
	u_0=\varepsilon=\pm 1,
\]
where $G_n$ is a polynomial in its arguments.  The first few terms in the series expansion are
\[
	\tilde w(\tilde z)
	=
	\frac\varepsilon{\zeta}-\frac{\varepsilon\tilde a_1(\tilde z_0)}6\zeta-\frac{\tilde a_0(\tilde z_0)+\varepsilon\tilde a_1'(\tilde z_0)}4\zeta^2
	+\left(\kappa+A(\tilde z_0)\log\zeta\right)\zeta^4+\cdots,
\]
where $A(\tilde z_0)=[2\tilde a_0'(\tilde z_0)+\varepsilon \tilde a_1''(\tilde z_0)]/2$.
In order to avoid the logarithmic singularity, we must have
\[
	2\tilde a_0'(\tilde z_0)+\varepsilon \tilde a_1''(\tilde z_0)=0
\]
for all $\tilde z_0\in\widetilde \Omega$ and for both $\varepsilon=1$ and $\varepsilon=-1$.  This implies the existence of constants $c_1$, $c_2$ and $c_3$ such that
$\tilde a_0(\tilde z)=c_1$ and $\tilde a_1(\tilde z)=c_2\tilde z+c_3$. Hence equation (\ref{new-pole-norm}\,b) becomes equation (\ref{almost-painleve}\,b).

Finally, we observe that equation (\ref{new-pole-norm}\,c) admits a solution of the form
\[
	\tilde w(\tilde z)=\sum_{n=0}^\infty u_n(x)\zeta^{(n-1)/2},
\]
where $u_n$ is a polynomial in $x=\log\zeta$ satisfying a recurrence relation of the form
\[
	(n+2)(n-6)u_n+4(n-2)\dot u_n+4\ddot u_n=H_n(u_0,\ldots,u_{n-1}),
	\qquad
	u_0=\eta,\quad \eta^4=1,
\]
where $H_n$ is a polynomial in its arguments.  The first few terms in the series expansion are
\begin{equation*}
\begin{split}
	\tilde w(\tilde z)
	&=
	\frac\eta{\zeta^{1/2}}
	-
	\frac{\tilde a_3(\tilde z_0)}4\eta^3\zeta^{1/2}
	-
	\frac{4\tilde a_2(\tilde z_0)}{15}\eta^2\zeta
\\
	&\quad +
	\frac{9\tilde a_3(\tilde z_0)^2-32[\tilde a_1(\tilde z_0)+\tilde a_3'(\tilde z_0)\eta^2]}{96}\eta\zeta^{3/2}
\\
	&\quad +
	\frac{6\tilde a_2(\tilde z_0)\tilde a_3(\tilde z_0)-20\tilde a_2'(\tilde z_0)\eta^2}{35}\zeta^2
	+
	\left(
		\kappa
		+
				A(\tilde z_0)\log\zeta
	\right)\eta\zeta^{5/2}+\cdots,
\end{split}
\end{equation*}
where $A(\tilde z_0)={\displaystyle\frac\eta 8}\left(2\tilde a'_1(\tilde z_0) - \tilde a_3(\tilde z_0)\tilde a'_3(\tilde z_0)+\eta^2\tilde a_3''(\tilde z_0)\right)$.
In order to avoid movable logarithmic branch points, we must have $A(\tilde z_0)=0$ for all $\tilde z_0\in\tilde\Omega$ and for $\eta^2=\pm 1$.  This gives
$\tilde a''_3(\tilde z)=0=\left[4\tilde a_1(\tilde z)-\tilde a^2_3(\tilde z)\right]'$.
Hence there are constants $c_1$, $c_2$ and $c_3$ such that $\tilde a_3(\tilde z)=2(c_2+c_3\tilde z)$ and 
$\tilde a_1(\tilde z)=(c_2+c_3\tilde z)^2+c_3$ and so equation (\ref{new-pole-norm}\,c) becomes equation (\ref{almost-painleve}\,c).
\hfill$\Box$

\section{Conclusion}
The Painlev\'e property is often used to detect integrable systems.  The algebro-Painlev\'e property is a generalisation that is invariant under algebraic changes of the dependent variables.  We have shown how to test for this property in several examples.  

\vskip 9 mm

\enlargethispage{20pt}

\noindent{\large \bf Acknowledgements}\\{The author thanks Maria Chivers for many fruitful discussions.}

\end{document}